\documentstyle[twocolumn,aps,floats]{revtex}

\input psfig.sty

\begin{document}
\draft

\twocolumn[\hsize\textwidth\columnwidth\hsize\csname %
@twocolumnfalse\endcsname



\title{Fluctuations in finite critical and turbulent systems}
\author{Vivek Aji and Nigel Goldenfeld}

\address{Department of
Physics, University of Illinois at Urbana-Champaign
\\1110 West Green Street\\
Urbana, IL, 61801-3080}

\date{\today}

\maketitle

\begin{abstract}

We show that hyperscaling and finite-size scaling imply that the
probability distribution of the order parameter in finite size critical
systems exhibit data collapse.  We consider the examples of equilibrium
critical systems, and a statistical model of ecology.  We explain
recent observations that the probability distribution of turbulent
power fluctuations in closed flows is the same as that of the harmonic 2DXY
model.

\end{abstract}
\vspace{0.2in}

]

Critical systems with identical symmetries, dimension and
exponents are defined to be members of the same universality
class; but must they also share the same probability distribution
for the fluctuating variables or order parameters?  And if two
systems do indeed exhibit the same probability distribution, are
they necessarily in the same universality class, in the
conventional meaning of the term?

Recently, light has been shed on these issues by studies of the
probability distribution functions (PDFs) of fluctuating quantities in
finite critical systems.  Bramwell, Holdsworth and Pinton (BHP)
\cite{BRAM} observed data collapse for the two-dimensional XY model
(2DXY) in the spin-wave regime at low temperatures and in statistical
models of nonequilibrium dynamics which exhibit self-organized
criticality (SOC). Even more remarkably, they found that the power
fluctuations in a closed turbulent flow
exhibit exactly the same form of data collapse, with a scaling function
that is indistinguishable from the aforementioned statistical critical
models.  Taken at face value, these observations lend support to the
notion that finite Reynolds number ($Re$) turbulence is indeed a critical
state, and that there is a kind of super-universality between systems
with different dynamics and even dimensionality.

The purpose of this Letter is to show that finite-size systems that are
in the critical regime should be expected to exhibit just this sort of
data collapse.  The system in question may be either an equilibrium
system near its critical point, or a nonequilibrium system that attains
a critical state through fine-tuning or other mechanism for achieving
scale invariance.  However, we are unaware of any reason {\it a
priori\/} to expect that the probability distribution should be
super-universal, and indeed we exhibit a counter example. Finally, we
argue that the apparent agreement between magnetic systems and
experiments on closed turbulent flows, while interesting and genuine,
is not indicative of the intrinsic behavior of turbulence; we propose
an explanation for the observations that appears to explain not only
the data collapse but the Reynolds number dependence as well.

{\it Probability Distribution Data Collapse:-}  Let us now review in
more detail the findings of BHP.  They examined the 2DXY model in the
harmonic approximation for temperatures well below the
Kosterlitz-Thouless transition $T_{KT}$.  Although in an infinite system the
magnetization $\vec{M}$ should be identically zero, for a finite system
there are large fluctuations, and they measured the probability
distribution function (PDF) for a range of system sizes and
temperatures.

The scaled PDFs of the magnetization below $T_{KT}$ collapse onto each
other for different system sizes and temperatures, provided one works
in the harmonic approximation and the correlation length,  $\xi$, is
larger than the system size. The scaling necessary to achieve this data
collapse is that the independent variable ($\eta\equiv |M|$) is
replaced by $y\equiv \eta - \left<\eta\right >/\sigma$, where $\sigma$
is a measure of the width of the PDF, such as the width at half
maximum. A similar data collapse was seen in a statistical model of
ecology \cite{HRT}, where the scaled probability distribution of
species abundance in a region was found to be independent of its area.
Pinton et al. \cite{PNT1,PNT2} performed experiments on confined turbulent
flows maintained at constant Reynolds number and looked at the PDF of
power fluctuations. These too showed data collapse across different
Reynolds numbers. Remarkably the PDFs for the turbulence data and the
2DXY model overlap within the precision of the data.

Subsequently a number of SOC systems, such as the Bak-Tang-Weisenfeld
sand pile model \cite{BTW}, the Sneppen depinning model \cite{SNP}, the
auto igniting forest fire model \cite{PSR} and a model for granular
media \cite{ECG}, have been studied and they too seem to show data
collapse with a PDF very similar to that observed in the turbulence
experiment \cite{BRAM1}. The same holds true of the 2D Ising and 2D
site percolation models as well. This seems to suggest that the scaling
form is independent of system attributes such as symmetry (discrete or
otherwise), state (equilibrium or otherwise) etc.

A similar phenomena was noted a long time ago by Nicolaides and Bruce
\cite{NB}, who were interested in the question of whether a
universality class is defined by the values of the critical exponents,
or whether the probability distributions were common to members of the
same universality class.  They found that the PDF of the two
dimensional Ising, spin $1$, and $\phi ^{4}$ models all had the same
form in finite systems.

In fact, the issue of data collapse is linked with the
existence of hyperscaling.  Let us begin with a discussion of data
collapse in equilibrium systems. In finite size magnetic systems, the
PDF has a scaling form in the critical region \cite{BND},
 $P_{L}(M) = L^{\beta/\nu}\widetilde{f}({L/\xi}, ML^{\beta/ \nu})$,
 where $\widetilde{f}$ is a scaling function, $\beta$, $\nu$ are the
critical exponents for the order parameter $M$ and the correlation
length $\xi$ respectively, and $L$ denotes the linear dimension of the
system, assumed to be in $d$ dimensions.

Near the critical point the correlation length becomes larger than the
system size and making the unjustified assumption (a priori) that
$\widetilde{f}$ is analytic in the first argument, we obtain
 $P_{L}(M) = L^{\beta / \nu}\widetilde{f}(ML^{\beta/ \nu}).$
The sum rule for the static susceptibility $\chi_{T}$ implies that
 $k_{B}T\chi_{T} \sim L^{\alpha}[\left<M^{2}\right> -
\left<M\right>^{2}]$.
Notice that the quantity in the brackets is the variance of the
probability distribution of the magnetization. The finite size scaling
form for the susceptibility is $\chi_{T} = L^{\gamma / \nu} F({L / \xi})$
In the limit $\chi/L$ goes to infinity the scaling function $F$ tends
towards a constant. Thus the measure of the width of the fluctuations,
$\sigma$, scales as $\sigma ^{2} \equiv \left<M^{2}\right> - \left<M\right>^{2}
\sim L^{{\gamma / \nu} - d}$

BHP found that $\sigma P(M/\sigma)$ is a universal function,
independent of system size. To test this, compute the function
 $\sigma P(M) = L^{{\gamma/ 2\nu} -d} L^{\beta / \nu}f(ML^{\beta / \nu})$,
\begin{equation}
\sigma P({M /\sigma}) = L^{(\gamma + 2\beta -d\nu) / 2\nu}f({M/
 \sigma}L^{{\beta/ \nu} +
 {\gamma/ 2\nu} - {d / 2}})
\end{equation}
Combining the hyperscaling relation, $2 - \alpha = d\nu$
and Rushbrooke scaling, $\alpha + 2\beta + \gamma = 2$
gives $\gamma + 2\beta = d\nu$.
Given this identity it follows that all $L$ dependence in ($1$)
disappears and we are left with a statement of data
collapse:
\begin{equation}
\sigma P({M/ \sigma}) = f({M/ \sigma})
\end{equation}
Thus as long as finite-size scaling holds true near the critical point and
hyperscaling is obeyed, the data, for different sizes, fall on top of
each other for a given system. To our knowledge, this was first
observed empirically by Nicolaides and Bruce \cite{NB}.

There are interesting consequences for the moments of the PDF, $P$,
illustrated here for the first two.  The mean of the distribution is
\begin{equation}
\left<M\right> = {\int MP(M)dM \over \int P(M)dM}\sim \sigma {\int zF(z)dz \over \int F(z)dz}
\end{equation}
where $F$ is a scaling function

The integral is a pure number, so that the ratio of the mean and the
variance is independent of $L$ ($\left<M\right> \sim \sigma$).

The moment relation is a direct result of hyperscaling,
 ${\left<M\right>} \sim {L^{\beta / \nu}}, {\sigma} \sim {L^{{\gamma / 2\nu} - {d / 2}}}$
\begin{equation}
{\left<M\right> \over \sigma} \sim {L^{-{2\beta +\gamma -d\nu / 2\nu}}}
\end{equation}
Hyperscaling then requires the $L$ dependence in the last line above to
vanish. This relation between the mean and variance has been explicitly
calculated for the $2$D XY in the spin wave approximation\cite{BRAM1}.
The moments of the order parameter satisfy,
${\mu_{n}} = {g_{n}(g_{2}/2)^{-n/2}\sigma^{n}}$ and
${\mu_{n}} \sim {\mu_{1}^{n} \sim \sigma^{n}}$ implying
\begin{equation}
\sigma \sim {\mu_{1} \sim \left<M\right>}
\end{equation}

{\it Probability Distribution Data Collapse in an Ecology Model:-} This
scaling is fundamental to data collapse and holds even for systems not
in thermal equilibrium. In particular it holds for the ecology model
mentioned earlier. Harte et al. \cite{HRT} proposed a model for the
species abundance distribution observed in nature. It has been
empirically observed that the number of species $S$ in a patch of area
$A$ obeys a scaling law,$S \sim A^{z}$.
Presumably this is an asymptotic in time form of a more general
dynamical model, as the time evolution within this model is not
specified. Nevertheless one can construct a recursion recursion
relation for the PDF, $P_{i}(n)$, the probability for finding a
species with $n$ individuals in a patch $i$.  Patches $i$ are
constructed by successive bifurcation of an initial biome.   The
key ingredient of the model is the assumption of self similarity, which
forces the species-area law, but allows the probability distribution for
the number of species resident in a biome to be calculated.
If a species is found in area $A_{i}$, then there is a non-zero
probability $a$, for finding it in one of the two halves (area
$A_{i+1}=A_{i}/2$,) which is independent of $i$.  The independence
on scale $i$ of the probability $a$ gives rise (or more
accurately, is equivalent) to the species-area rule, with $a =
2^{-z}$. The resulting equation turns out to be
\begin{equation}
P_{i}(n) = xP_{i+1}(n) + (1-x)\sum_{k=1}^{n-1}P_{i+1}(n-k)P_{i+1}(k)
\end{equation}
where $x = 2(1 - a)$. The variance of this PDF was computed by Banavar et
al. \cite{BNV}:
\begin{equation}
\sigma_{i}^{2} = \sum_{j=0}^{m-i-1} (2-x)^{j}(2-x)^{m-i-1}x(1-x)
\end{equation}
where $m$ is defined as the maximum number of times the
system can be subdivided before no species exist on the smallest patch
($P_{m}(1)=1$). When the system size is much bigger than the size of
this smallest patch, the variance obeys a simple recursion relation,
${\sigma_{i} / \sigma_{i+1}} \sim 2-x$.
The data collapse observed for this model is the statement
${\sigma_{i}P_{i}(n)} = {\sigma_{i+1}P_{i+1}(n')}$ and
${n/\sigma_{i}} = {n'/\sigma_{i+1}}$.
It can be shown\cite{MG} that the PDF satisfies a finite-size scaling
relation of the form $P_{i}(n) = {1/n}f({n/ N_{i}^{\phi}})$,
where $\phi$ is a crossover scaling exponent, $N_{i}$ is the
number of individuals of all species in biome patch $i$, and
$N_{i+1}^{\phi}
 = 2^{-\phi}N_{i}^{\phi}$; then
the data collapse is equivalent to a relation between the
exponents.
\begin{equation}
{f({n \over N_{i}^{\phi}})} = {f({n \over (2-x)N_{i}^{\phi}2^{-\phi}})}
\end{equation}
leading to $ 2-x = {2^{\phi}}$ which is equivalent to $\phi + z = 1$.

This relation is nothing but the hyperscaling relation observed in
the magnetic system. This can be seen by considering the
asymptotic forms of the mean and variance of the PDF,
${<n>} \sim  {N^{\phi}}$ and $ {\sigma} \sim  {(2-x)^{m} \sim 2^{m\phi}} \sim  {N^{\phi}}$.
This immediately gives us the moment scaling which was crucial to data collapse
and hyperscaling, $\sigma \sim \left<n\right>$.
It is remarkable that the ecology model shows this behaviour for its
moments -- an unforseen consequence of the assumption of
self-similarity.

{\it Power fluctuations in a turbulent flow:-}  Having established the
connection between hyperscaling (moment relationships) and data
collapse let us turn our attention to the case of confined turbulent
flows. The experiment consists of a closed cylinder in which a
turbulent fluid is driven at the top and bottom by counter-rotating
plates with vanes, moving at the same mean angular frequency.  The PDFs
for power fluctuations $\Pi$ were measured \cite{PNT1,PNT2} for different
Reynolds numbers ($Re$) and found to exhibit data collapse. However the
ratio of the mean to the variance depended on $Re$:
$\Pi_{rms}/\overline{\Pi} \sim Re^{-\alpha},\hskip 0.5cm  \alpha
\sim 0.33$.

This observation shows that the Reynolds number, and hence the
extent of the inertial range, is not the parameter that controls
the nature of the fluctuations: it is not analogous to the finite
size of the system, as has been suggested previously.  If this
suggestion were correct, hyperscaling would have required that the
ratio of mean to variance be independent of Reynolds number.

Our analysis suggests that the relevant parameter, in the case of
the confined flows, that controls the nature of the fluctuations
is not the Reynolds number but the physical system size itself. In
addition to the scaling analysis mentioned above, the reasons for
this is two fold. First, the experiments were performed by
changing either the fluid (i.e. viscosity) or the angular velocity
(see Ref. 3 and 4 for details of the experiment) and not by changing the
spatial dimensions of the flow. Second, it was also noted that an
open flow did not exhibit any non-Gaussian fluctuations in the total power
dissipated.
In other words the fluctuations are not governed by the degrees of the
turbulent flow (i.e. Reynolds number) but the degrees of freedom of
coherent structures that form in these system which in turn depends on
the system size, $L$.

We hypothesize that the flow is composed of a turbulent region around
the top plate with a mean angular momentum, another oppositely directed
turbulent region around the bottom plate, and an interfacial region
between them -- a shear pancake.  Experiments were performed in both
open and closed geometries, which differ by the absence or presence of
a confining cylindrical wall.  In the open geometry, the power
fluctuations of each plate were non-Gaussian, and negatively skewed,
while the total power (sum of the power measured at each plate)
fluctuations were Gaussian.  In the closed geometry all of these three
quantities were non-Gaussian, with a scaled probability distribution
apparently indistinguishable from that of the 2D XY model simulations.

In the open geometry, the shear pancake can shift its mean vertical
position instantaneously as shear energy is dissipated horizontally,
but in the closed geometry, this is not possible because of the walls.
In the open geometry, as the shear pancake moves, the net turbulent
energy in the upper half of the cell will increase/decrease while that
in the lower half of the cell will decrease/increase: hence, we
anticipate that the power fluctuations should be anti-correlated, in
agreement with observations\cite{labbefigs}.

In the closed geometry, we can describe the shear pancake by its height
$h(x,y)$ above the $x$, $y$ plane positioned parallel to and
equidistant from the rotating plates.  The shear pancake experiences
random fluctuations from the turbulent flows, which drive it with an
effective, Reynolds number dependent temperature $T(Re)$ and Boltzmann
distribution $P_h(h) \propto \exp[-S(h)/T]$.  The action $S(h)$
describes local height fluctuations of the shear pancake, and the power
dissipation of the system should be expected to depend in some way on
the friction between the two counter-rotating turbulent flows i.e. on
the surface area of the shear pancake, and not its absolute mean height.
Any local fluctuation increases the total surface area of the shear
pancake and hence increases the power dissipation in both flows.
Hence, in this case, we anticipate that the power fluctuations would be
correlated, in agreement with observations\cite{labbefig4}.  Thus the
action should be of the form
\begin{equation}
S(h) \propto \int dx\,dy\,\sqrt{1+(\nabla h)^2} \approx \int dx\,dy\,(\nabla h)^2
\end{equation}
which is the Hamiltonian of the 2D XY model in the spin wave
approximation, making the identification of $h$ with the phase
$\theta$.

In order to complete the dictionary between the PDFs of the 2D XY model
magnetization and the turbulent power, we note that the power
dissipated is by hypothesis proportional to $S\{h\}$ and thus is linearly
dependent on $(\nabla h)^2$.  Similarly, the absolute value of the
magnetization per spin $M$, measured in the 2DXY simulations (for $N$ spins
$\vec S_i$) is given by
$
M\equiv \sqrt{\left(\sum_{i=1}^N \vec S_i\right)\cdot
\left(\sum_{j=1}^N \vec S_j\right)}/N\approx 1 -
\frac{1}{4N}\sum_{ij}(\theta_i-\theta_j)^2,
$
showing that the probability distribution of magnetization fluctuations is indeed the
counterpart of the power fluctuations in the turbulent flow.  It
follows from our assumptions above that the probability distributions
should be identical.  Note that we are definitely predicting that the
power probability distribution should be that of the 2DXY model in the
harmonic approximation only.  Indeed, we have simulated the 2DXY model
for temperatures near the Kosterlitz-Thouless transition, and find
significant deviations from the probability distribution for the
turbulent power and the 2DXY model in the harmonic approximation as shown in
figure 1 (and has also been noticed by Palme et al. \cite{PML}).

\begin{figure}[h]
\leavevmode\centering\psfig{file=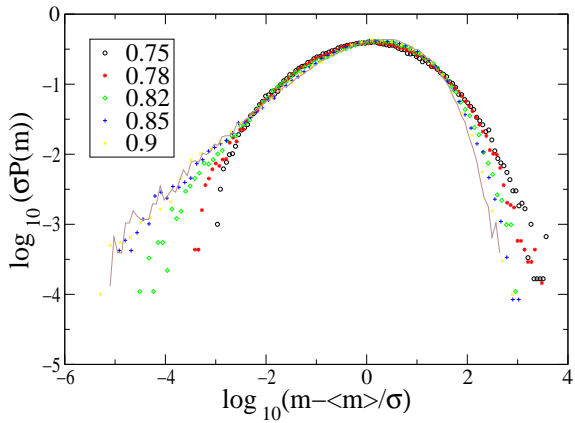,width=\columnwidth,angle=-90}
\caption{Probability distribution function of the magnetization
in a 2DXY model for different values of temperature: $ J/K_{B}T = 0.75(\circ),
0.78(*), 0.82(\diamond), 0.85(+)$ and $0.9(\times)$. The solid line represents
the universal distribution of BHP [1].}
\end{figure}

Now, we discuss the Reynolds number dependence of the ratio
$\mu/\sigma$.  By hypothesis, this should be given by using the
temperature dependence of $\mu/\sigma$ from the harmonic 2D XY model,
but with $T$ given by $T(Re)$.  To see how the effective temperature
should scale with $Re$, note that the velocity of the flow scales as
$R\Omega$ where $R$ is the radius of the plates, and $\Omega$ is the
angular velocity. The net mass per unit height of the cylinder scales
as $R^2$. Hence the net kinetic energy scales as $R^4\Omega^2$. The
$Re$ number is proportional to $R^2\Omega$ so that the energy scales as
$Re^2$.  The number of degrees of freedom giving rise to this turbulent
energy is proportional to $Re^{9/4}$, so that $T(Re) \sim Re^{-1/4}$.
Using the fact that $\mu/\sigma\sim T$ \cite{ARCH}
we obtain that $\mu/\sigma\sim Re^{-1/4}$, which agrees reasonably
with the data, although the exponent is not the optimal fit to all the
data points, as shown in figure 2.
\begin{figure}[htb]
\leavevmode\centering\psfig{file=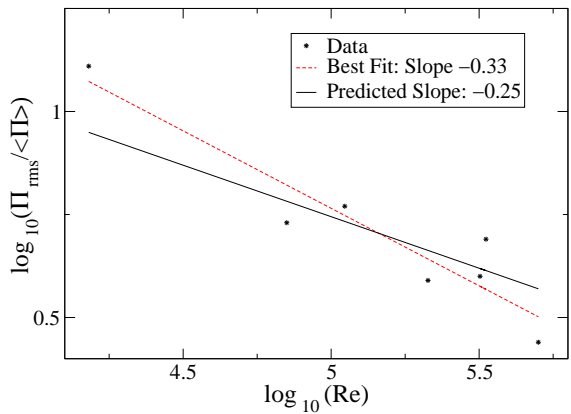,width=\columnwidth,angle=-90}
\caption{Data from J-F Pinton et al. [4] with the best fit and
a fit to our predicted slope of -$0.25$.}
\end{figure}

Finally, we mention that the dynamic universality class of the height
fluctuations should be the 2D Edwards-Wilkinson model \cite{EW}; implications of this for the time
dependent correlations of the power will be discussed elsewhere.

In conclusion, we have shown how universal scaling phenomena can arise
in finite critical systems due to hyperscaling; the observed similarity of the
probability distribution scaling function in the harmonic 2DXY
and a closed turbulent cell is a special case, and is not generic.

\acknowledgements We acknowledge useful discussions with
E. Bodenschatz, J.-F. Pinton and Z. R\'{a}cz and other participants of the
workshop on Universal Fluctuations in Correlated Systems, held at
Lyons, April 2000, where some of this work was reported.  This work was
supported by National Science Foundation grant NSF-DMR-99-70690.


\begin{thebibliography}{99}
\bibitem{BRAM}
S.T. Bramwell, P.C.W. Holdsworth and J.-F. Pinton, Nature {\bf 396},
552 (1999).

\bibitem{HRT}
J. Harte, A. Kinzig and J.L. Green, Science {\bf 284}, 334 (1999).

\bibitem{PNT1}
R. Labbe, J.-F. Pinton and S. Fauve, J. Phys. II France {\bf 6}, 1099
(1996).

\bibitem{PNT2}
J.-F. Pinton and P.C.W. Holdsworth, Phys. Rev E {\bf 60}, R2452
(1998).

\bibitem{BTW}
P. Bak, C. Tang and K. Weisenfeld, Phys. Rev. Lett. {\bf 59}, 381
(1997).

\bibitem{SNP}
K. Sneppen, Phys. Rev. Lett. {\bf 69}, 3538 (1992).

\bibitem{PSR}
P. Sinha-Ray, L.B. de Agua and H.J. Jensen, to be published.

\bibitem{ECG}
E. Caglioti, V. Loreto, H. Hermann and M. Nicodemi, Phys. Rev. Lett.
{\bf 79}, 1575 (1997).

\bibitem{BRAM1}
S.T. Bramwell et al., Phys. Rev. Lett {\bf 84}, 3744 (2000). For other examples
of order parameter fluctuations in finite systems see R. Botet and
M. Ploszajczak, unpublished.


\bibitem{NB}
D. Nicolaides and A.D. Bruce, J. Phys. A:Math. Gen. {\bf 21}, 233 (1988).

\bibitem{BND}
K. Binder, {\it Computational Methods in Field Theory}, edited by H.
Gauslever and C.B. Lamb (Springer, Berlin, 1992).

\bibitem{BNV}
J.R. Banavar, J.L. Green, J. Harte and A. Martin, Phys. Rev. Lett. {\bf 83},
4214 (1999).

\bibitem{MG}
H.G. Martin and Nigel Goldenfeld, unpublished.

\bibitem{labbefigs}
See the ref. \cite{PNT2}, figure 2(b).

\bibitem{labbefig4}
See the ref. \cite{PNT2}, figure 4(b).

\bibitem{PML}
G. Palma, T. Meyer and R. Labbe, unpublished.

\bibitem{ARCH}
P. Archambault, S.T. Bramwell and P.C.W. Holdsworth, J.Phys. A Math. Gen. {\bf 30},
8363 (1997).

\bibitem{EW}
S.F. Edwards and D.R. Wilkinson, Proc. Roy. Soc. A {\bf 38}, 17 (1982).

\end{thebibliography}
\end{document}